\documentclass[12pt]{iopart}
\usepackage{graphicx}
\usepackage{iopams}
\usepackage{subcaption}
\usepackage{color}
\begin{document}

\title{Miscible-Immiscible Transition and Nonequilibrium Scaling in Two-Component Driven Open Condensate Wires}

\author{Liang He$^{1,2}$ and Sebastian Diehl$^{1,2}$}

\address{$^1$Institute for Theoretical Physics, University of Cologne, D-50937
Cologne, Germany}

\address{$^2$Institute for Theoretical Physics, Technical University Dresden,
D-01062 Dresden, Germany}

\begin{abstract}
We investigate the steady state phase diagram of two-component
driven open condensates in one dimension. We identify a miscible-immiscible transition
which is predominantly driven by gapped density fluctuations and
occurs upon increasing the inter-component dissipative coupling. Below the transition in the miscible phase, we find the effective long wavelength
dynamics to be described by a two-component Kardar-Parisi-Zhang (KPZ)
equation that belongs to the nonequilibrium universality class of
the one-dimensional single-component KPZ equation at generic choices
of parameters. Our results are relevant for different experimental
realizations for two-component driven open condensates in exciton-polariton
systems. 
\end{abstract}

\section{Introduction}
Recent experimental development in nanoscience, quantum optics, ultracold
atom physics and related areas has given rise to new classes of synthetic
physical systems with properties of both fundamental and practical
interests. One class of these systems are quantum many body ensembles
in a driven open setting, which includes exciton-polaritons in semiconductor
heterostructures~\cite{Kasprzak_nature_2006,Lagoudakis_NatPhys_2008,Love_PRL_2008,Roumpos_PNAS_2012,Bloch_PRL_2012,Takemura_NatPhys_2014,Fischer_PRL_2014,Ohadi_PRX_2015,Askitopoulos_PRB_2016},
ultracold atoms~\cite{Syassen_Science_2008,Carr_PRL_2013,Zhu_PRL_2014},
trapped ions~\cite{Blatt_NatPhys_2012,Britton_Natrue_2012}, and
microcavity arrays~\cite{Hartmann_laser_photonics_rev_2008,Houck_NatPhys_2012}.
The unifying and characteristic trait of these systems is the breaking
of detailed balance on the microscopic level, making them promising
laboratories to advance the frontier of nonequilibrium statistical
physics.

One current research focus are driven open condensates (DOCs). They
can be realized, for instance, in exciton-polariton systems~\cite{Kasprzak_nature_2006,Lagoudakis_NatPhys_2008,Love_PRL_2008,Roumpos_PNAS_2012,Bloch_PRL_2012,Takemura_NatPhys_2014,Fischer_PRL_2014,Ohadi_PRX_2015,Askitopoulos_PRB_2016}.
On the one hand, recent theoretical investigations~\cite{Altman_PRX_2015,Gladilin_PRA_2014,Ji_PRB_2015,He_PRB_2015, Mathey_PRA_2015,Wachtel_PRB_2016,Sieberber_PRB_2016,He_PRL_2017}
on single-component DOCs have revealed a new class of nonequilibrium
phenomena, that are related to a variant of the famous Kardar-Parisi-Zhang
(KPZ) equation~\cite{KPZ_PRL_1986} in nonequilibrium statistical
physics. On the other hand, investigations on multi-component
Bose-Einstein condensates (BECs), realized mainly in ultracold atom
experiments, have shown that the inter-component couplings generally
give rise to very rich physics~\cite[and the references therein]{Stamper-Kurn_RMP_2013}.
For instance, in the simplest case, i.e., two-component BECs, it is
well known that a strong enough inter-component interaction drives
a miscible-immiscible transition, also called phase separation, via fluctuations of sound mode type~\cite{Myatt_PRL_1997,Hall_PRL_1998,Nicklas_PRL_2011,Timmermans_PRL_1998}, and where, in the phase separation the two components avoid each other spatially.
Its existence can thus be traced back to the closed nature of these
systems, namely, to the simultaneous presence of both particle number
and momentum conservation. These are ingredients that are explicitly
violated in their driven-open counterparts. Such a setting therefore
gives rise to a novel scenario for the miscible-immiscible transition,
and, more generally, for the physical effects of inter-component couplings
in multi-component condensates in a generic driven-open setup. In
light of recent developments realizing multicomponent DOCs via different
polarization degrees of freedom~\cite{Fischer_PRL_2014,Ohadi_PRX_2015,Askitopoulos_PRB_2016}
or polaritonic Feshbach resonances~\cite{Takemura_NatPhys_2014},
understanding the generic features of the phase diagram of such systems
becomes a pressing task.

\begin{figure}
\begin{subfigure}{6in}
 \centering
\includegraphics[height=2in]{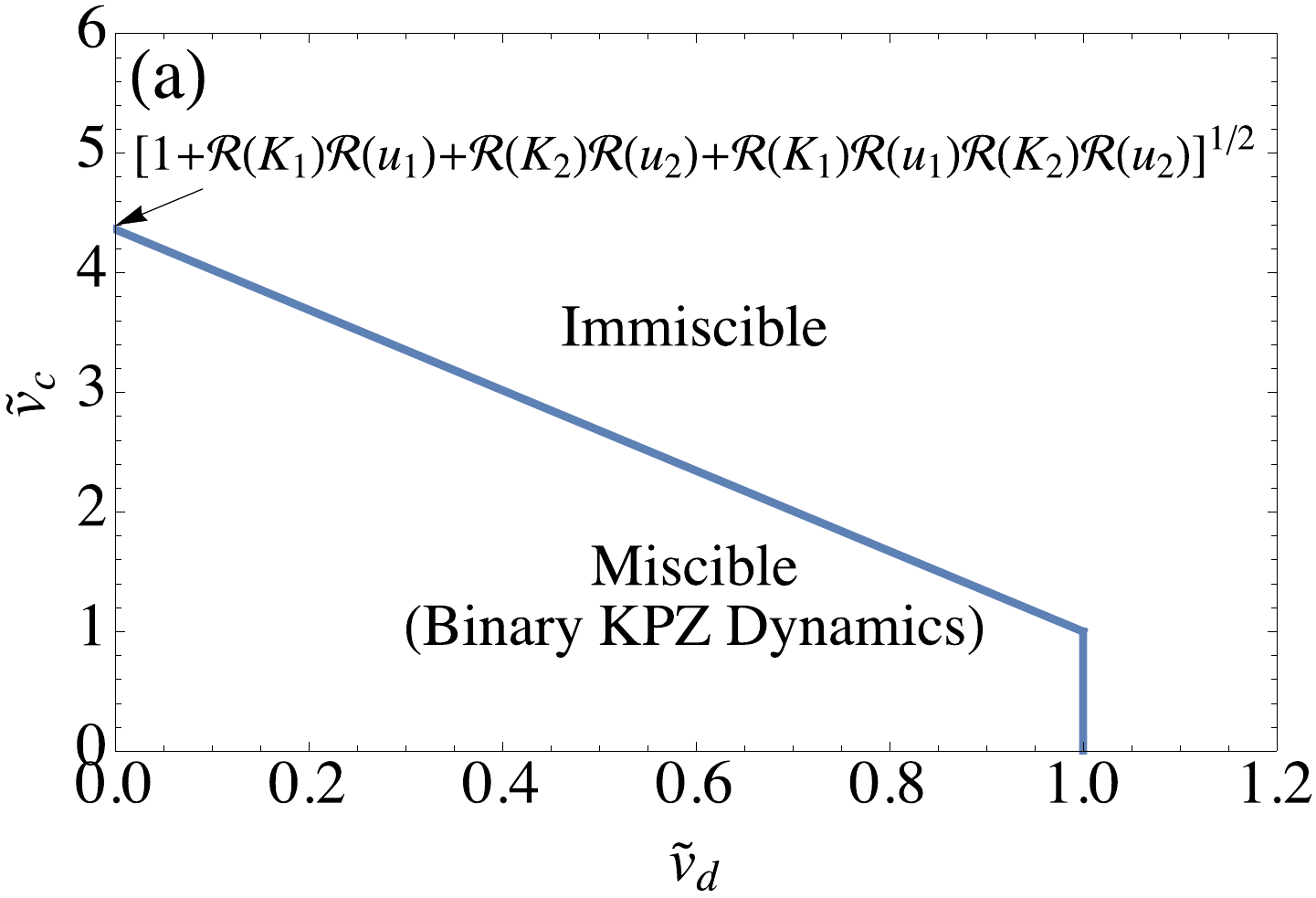}
\end{subfigure}

\begin{subfigure}{6in}
\includegraphics[height=1.5in]{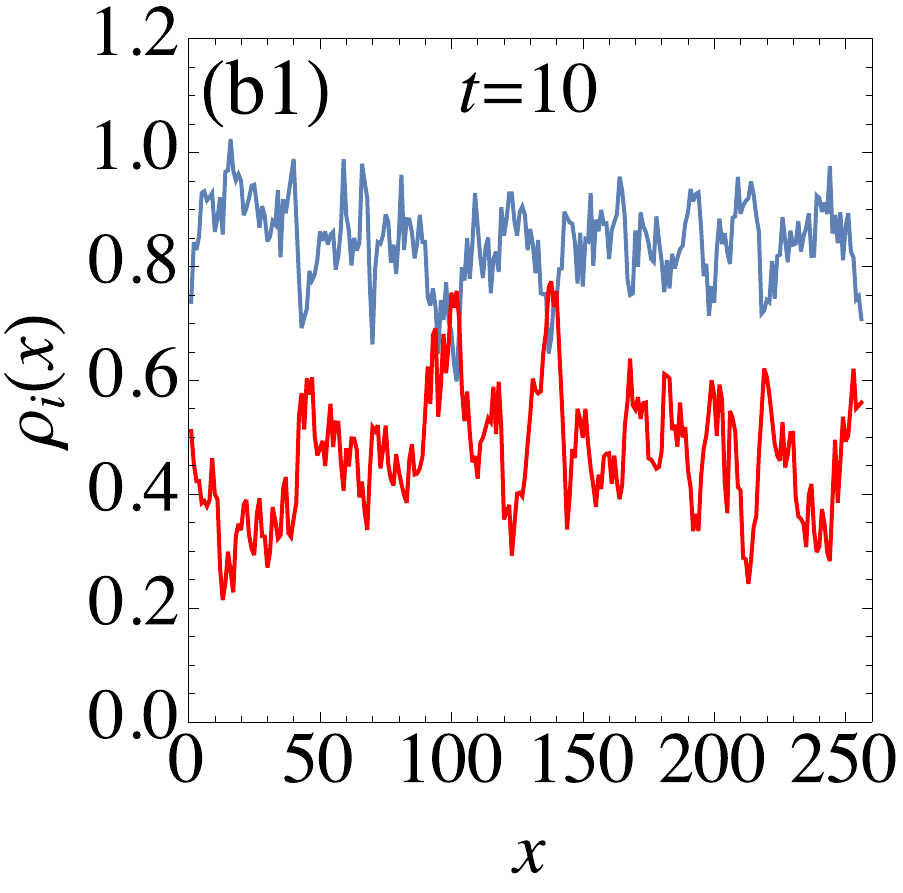}\includegraphics[height=1.5in]{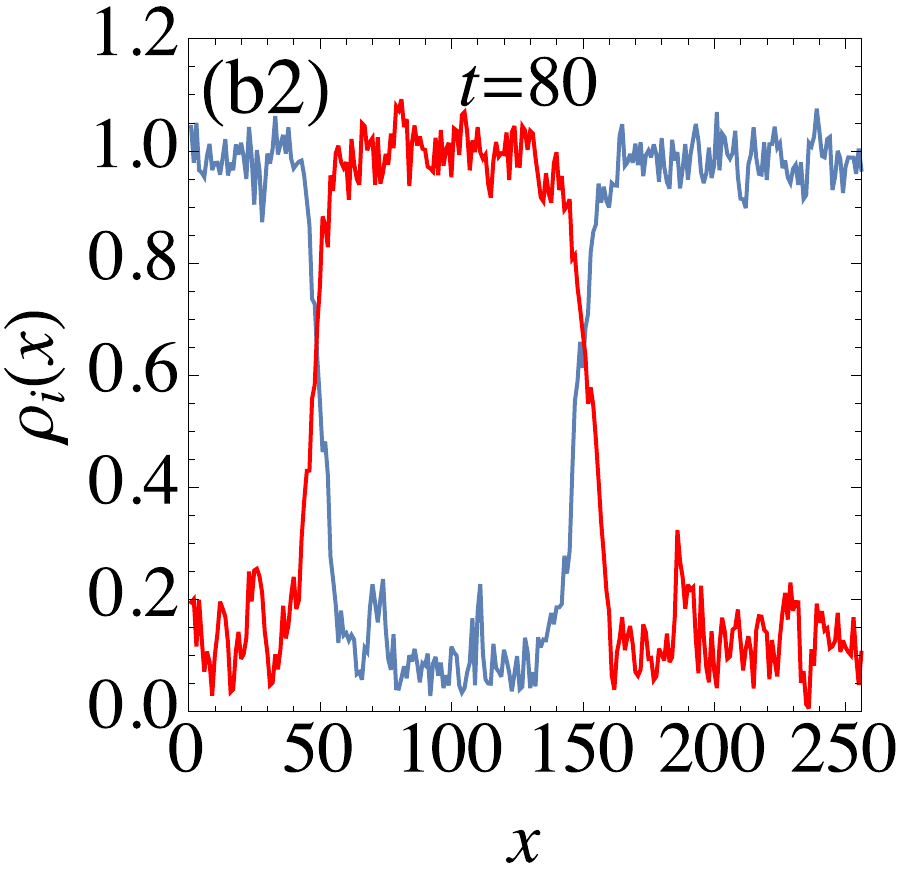}\includegraphics[height=1.5in]{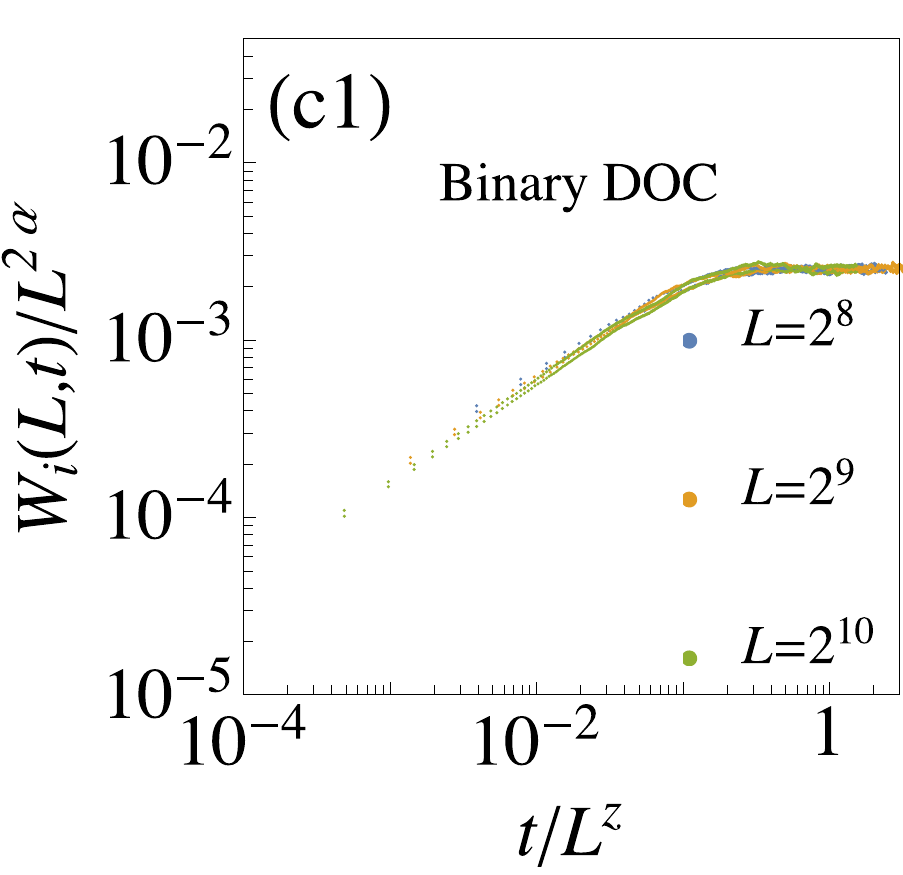}\includegraphics[height=1.5in]{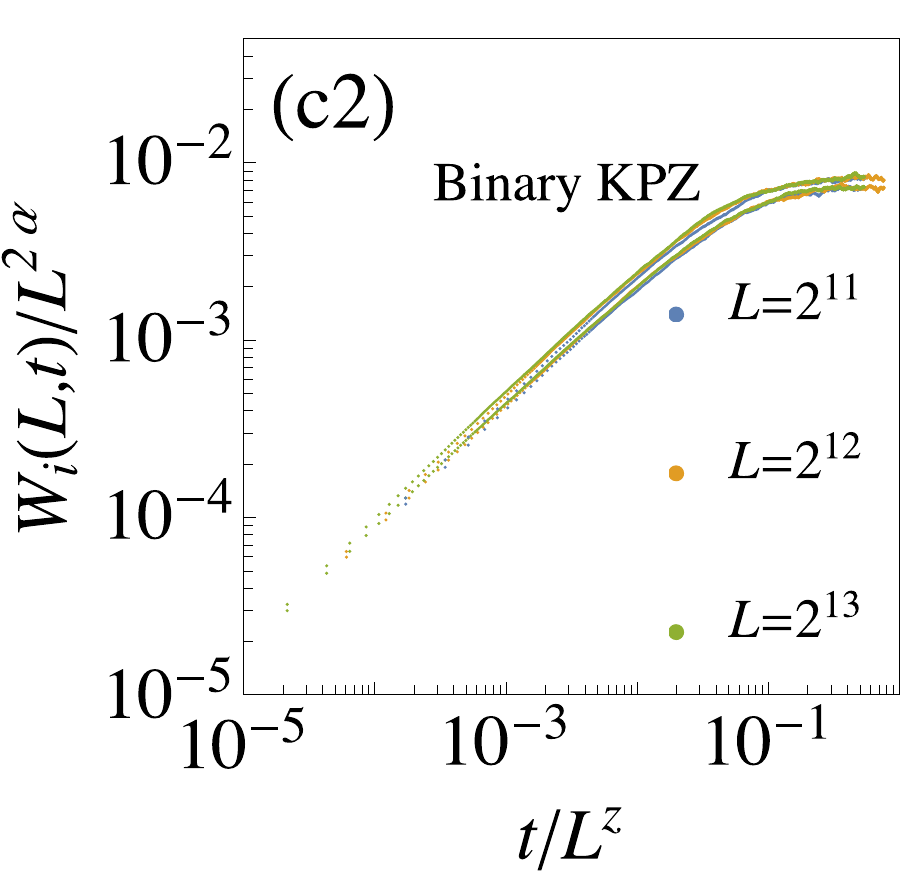}
\end{subfigure}

\caption{(a) Phase diagram of the system in terms of the rescaled dimensionless
inter-component coherent coupling $\tilde{\upsilon}_{c}$ and the
dissipative one $\tilde{\upsilon}_{d}$ at weak noise level and under
weak nonequilibrium condition.  Function
$\mathcal{R}(z)\equiv\Re(z)/\Im(z)$ for $z\in\mathbb{C}$. The values of other parameters used
in the phase digram are $r_{1,d}=u_{1,d}=K_{1,d}=1$, $r_{1,c}=u_{1,c}=0.1$,
$K_{1,c}=3.0$; $r_{2,d}=u_{2,d}=1.1,\, K_{2,d}=1.05$, $r_{2,c}=u_{2,c}=0.11$,
$K_{2,c}=3.1$. (b1, b2) Snapshots of distributions of condensate
field amplitudes $\rho_{j}(x)\equiv|\psi_{j}(x)|$ in the immiscible
phase at different time (the blue and red curve correspond to $\rho_{1}(x)$
and $\rho_{2}(x)$, respectively), where an immiscible phase is clearly
observed at a sufficiently long time ($t=80$). The values of other parameters
are $\sigma_{i}=0.01$, $v_{d}=1.2$, $v_{c}=0.05$. (c1, c2) Finite
size critical scaling collapse for the roughness function $W_{j}(L,t)$
for the two-component DOC in the miscible phase (c1) and for the two-component
KPZ equation Eq.~(\ref{eq:2-component-KPZ}) (c2). For the two-component
DOC, the value of other parameters are $\sigma_{1}=0.04,\sigma_{2}=0.05$,
$v_{d}=0.6$, $v_{c}=0.2$. For the two-component KPZ equation, the
parameters are $(D_{11},D_{12},D_{21},D_{22})=(1,0.5,0.55,1.1)$,
$(\lambda_{11},\lambda_{12},\lambda_{21}\lambda_{22})=(7,2,2.5,7.5)$,
$(\sigma_{1}^{\mathrm{KPZ}},\sigma_{2}^{\mathrm{KPZ}})=\left(0.1,0.11\right)$.
The lower and upper set of curves in each plot correspond to $W_{1}(L,t)$
and $W_{2}(L,t)$, respectively. The critical exponents of the 1D
\emph{single-component} KPZ equation $z=3/2,\,\alpha=1/2$ yield a
good finite size data scaling collapse shown in (c1, c2), which suggests
that the dynamics of two-component DOC in the miscible phase
and the two-component KPZ equation belong to the same dynamical universality
class of the\emph{ }1D\emph{ single-component} KPZ equation at generic
choices of parameters. See text for more details.}

\label{Fig. Phase_diagram} 
\end{figure}

As a first step to address questions along this line, we investigate
one-dimensional (1D) two-component DOCs at relatively weak noise strength
and under weak nonequilibrium condition in this work, with particular
focus on the physical effects of the inter-component couplings. 
To this end, we establish the phase diagram of the system in terms of the dimensionless inter-component coherent coupling (elastic two-body collisions) $\tilde{v}_{c}\equiv v_{c}/\sqrt{u_{1,c}u_{2,c}}$ and dissipative coupling (two-body losses) $\tilde{v}_{d}\equiv v_{d}/\sqrt{u_{1,d}u_{2,d}}$ as shown in Fig.~\ref{Fig. Phase_diagram}(a), where $v_{c}$ and $v_{d}$ are the coherent  and dissipative  inter-component couplings, respectively; $u_{j,c}$ and $u_{j,d}$ are the intra-component elastic collision strength and positive two-particle loss rate, respectively, with $j=1,2$ being the component index.
More specifically, we find: (i) A miscible-immiscible transition at finite $\tilde{v}_{d}$. 
Interestingly, the mechanism behind that transition is different depending on the regime of $\tilde{v}_{d}$. 
For $\tilde{v}_{d}>1$, the transition is driven by $\tilde{v}_{d}$ itself via the gapped density fluctuations [cf.~Fig.~\ref{Fig. Phase_diagram}(a) and Eq.~(\ref{eq:omega_1_3})]. We notice that this condition $\tilde{v}_{d}>1$ interestingly shares a form analogous to the one in the purely coherent two-component BEC, i.e., $\tilde{v}_{c}>1$~\cite{Timmermans_PRL_1998}. The mechanism underlying the transition is however vastly different, since in purely coherent two-component BECs, the transition is driven by the gapless sound modes~\cite{Timmermans_PRL_1998}. For $\tilde{v}_{d}<1$, the transition is
driven by the inter-component coherent coupling $\tilde{v}_{c}$ via
gapless \emph{diffusive} modes {[}cf.~Eq.~(\ref{eq:omega_2_4}){]},
whose critical value can be strongly affected by the presence of dissipative
couplings in the system {[}cf.~Fig.~\ref{Fig. Phase_diagram}(a),
Eqs.~(\ref{eq:PS_condition_leading_order}), (\ref{eq:PS_condition_leading_order_at_zero_vd}){]}. The clearest signatures of the immiscible state is present in single experimental runs, where the two components avoid each other spatially
(ii) Below the transition in the miscible phase, we find that the long
wavelength dynamical behavior of the system is effectively described
by a two-component KPZ equation, where, in particular, the KPZ-nonlinearity
of one component is coupled to the dynamics of the other one {[}cf.~Eq.~(\ref{eq:2-component-KPZ}){]}.
We further show that this two-component KPZ equation, and hence also the
dynamical behavior of two-component DOCs, belongs to the nonequilibrium
dynamical universality class of the 1D \emph{single-component} KPZ
equation at generic choices of parameters {[}cf.~Figs.~\ref{Fig. Phase_diagram}(c1,
c2){]}. In other words, at generic choices of parameters, the inter-component
coupling is irrelevant in the renormalization group (RG) sense in
the miscible phase, highlighting the remarkable degree of universality of KPZ physics.

\section{Microscopic model}

We describe the dynamics of the two-component 1D DOC by a generic
minimal model that assumes the form of a two-component version of
the stochastic complex Ginzburg-Landau equation (SCGLE) as appropriate
for the exciton-polariton systems~\cite{Ohadi_PRX_2015,Askitopoulos_PRB_2016,Altman_PRX_2015,Wouters_PRL_2007,Carusotto_RMP_2013,Pinsker_PRL_2014,Xu_arXiv_2017},
which reads (units $\hbar=1$)
\begin{equation}
\partial_{t}\psi_{j}=\left(K_{j}\partial_{x}^{2}+r_{j}-u_{j}|\psi_{j}|^{2}-v|\psi_{\bar{j}}|^{2}\right)\psi_{j}+\zeta_{j},\label{eq:EOM_2c_doc}
\end{equation}
with indices $j,\bar{j}=1,2$ and $\bar{j}\neq j$, denoting different
components. In exciton-polariton systems, they correspond to the two
polarization directions of polaritons~\cite{Takemura_NatPhys_2014,Fischer_PRL_2014}.
Here $K_{j}=K_{j,d}+iK_{j,c}$, $r_{j}=r_{j,d}+ir_{j,c}$, $u_{j}=u_{j,d}+iu_{j,c}$,
$v=v_{d}+iv_{c}$, $\langle\zeta_{j}^{*}(x,t)\zeta_{j'}(x',t')\rangle=2\sigma_{j}\delta_{jj'}\delta(x-x')\delta(t-t')$
and $\langle\zeta_{j}(x,t)\zeta_{j'}(x',t')\rangle=0$, where $K_{j,c}=1/m_{\mathrm{LP}}$
with $m_{\mathrm{LP}}$ being the effective polariton mass in exciton-polariton
systems, and $K_{j,d}$ is a diffusion constant. $r_{j,c}$ simply
reflects the choice of the rotating frame and can be modified by changing
to a different frame, i.e., $\psi_{j}(x,t)=\psi_{j}'(x,t)e^{i\mu_{j}t}$,
with modified $r_{j,c}'=r_{j,c}+\mu_{j}$. $u_{j,c}$ is the elastic
collision strength. $r_{j,d}=\gamma_{j,p}-\gamma_{j,l}$ is the difference
between the single particle pump and loss, denoted by $\gamma_{j,p}$
and $\gamma_{j,l}$, respectively. For the existence of condensates
in the mean field steady state solution, $r_{j,d}$ has to be positive,
i.e., the single-particle pump rate has to be larger than the loss
rate. $u_{j,d}$ is the positive two-particle loss rate. $v_{c}$
and $v_{d}$ are the coherent (elastic) and dissipative
inter-component couplings, respectively. Both of them are assumed
to be positive in the following, indicating a positive inter-component
elastic collision strength and positive inter-component two-particle
loss rate. In exciton-polariton systems, $u_{j,d}$ and $v_{d}$ originate
from intra-component and inter-component gain-saturation nonlinearities~\cite{Ohadi_PRX_2015},
respectively, instead of additional loss mechanisms. Consequently
the noise strength $\sigma_{j}$ is set by the single particle loss
$\gamma_{j,l}$, i.e., $\sigma_{j}=\gamma_{j,l}$ (cf.~\cite{Altman_PRX_2015}
for a related discussion in the single component case). We obtain
most of the numerical results presented in this work by directly solving
Eq.~(\ref{eq:EOM_2c_doc}) using the same numerical approach as in
Ref.~\cite{He_PRB_2015} and set $r_{1,d}=K_{1,d}=1$, indicating
$t$ and $x$ are measured in units of $r_{1,d}^{-1}$ and $\sqrt{K_{1,d}}$,
respectively. For performing the ensemble average, we use $10^{3}$
stochastic trajectories if not mentioned otherwise.

\section{Miscible-immiscible transition}

In the context of ultracold atom physics, it is well known that in
multi-component BEC systems, large enough inter-component interactions
can drive various miscible-immiscible transitions~\cite{Myatt_PRL_1997,Hall_PRL_1998,Nicklas_PRL_2011,Timmermans_PRL_1998,Kasamatsu_PRL_2004,He_PRA_2009,Vidanovic_NJP_2013}.
In particular, in two-component BECs, if the inter-component interaction
$v_{c}>\sqrt{u_{1,c}u_{2,c}}$, it can drive a transition from a miscible
phase to an immiscible phase or phase separation~\cite{Myatt_PRL_1997,Hall_PRL_1998,Nicklas_PRL_2011,Timmermans_PRL_1998},
where particles of different components stay away from each other
in space. The dynamical reason that drives this transition comes from
the instability caused by fluctuations of the sound mode type~\cite{Timmermans_PRL_1998},
whose existence relies on the condition that both the particle number
and the momentum conservation are present. However, for two-component
DOCs, this condition is apparently absent due to the driven-open characteristic
of the system. More interestingly, the inter-component \emph{dissipative}
coupling $v_{d}$ is clearly a new relevant coupling in two-component
DOCs that could be important in the miscible-immiscible transition.
In the following, we shall investigate the key factors that determine
the miscible-immiscible transition in this driven-open case and identify
the dynamical modes that are responsible for that transition.

To this end, we perform a leading order stability analysis on the
system's deterministic dynamics, where we choose $r_{j,c}$ in such
a way that, in the absence of noise, the equation of motion (EOM)
Eq.~(\ref{eq:EOM_2c_doc}) has a stationary, spatially uniform solution,
and linearize the system's deterministic dynamics around this solution.
We denote the stationary, spatially uniform solution as $\psi_{j}^{(0)}(x,t)$,
whose explicit form reads $\psi_{j}^{(0)}(x,t)=M_{j},$ with $M_{j}$
being the amplitude of $\psi_{j}^{(0)}$ that solves the homogenous
real part of Eq.~(\ref{eq:EOM_2c_doc}) with the explicit form $M_{j}^{2}=(r_{j,d}u_{\bar{j},d}-r_{\bar{j},d}v_{d})/(u_{1,d}u_{2,d}-v_{d}^{2})$.
In general, $\psi_{j}$ can be expressed as the sum of $\psi_{j}^{(0)}$
and fluctuations $\delta\psi_{j}$ on its top, i.e., $\psi_{j}=\psi_{j}^{(0)}+\delta\psi_{j}$,
where fluctuation fields $\delta\psi_{j}$ can be further decomposed
into their Fourier components, i.e., $\delta\psi_{j}(x,t)=\sum_{k}c_{j,k}(t)\exp(ikx)$,
with $c_{j,k}(t)$ being fluctuation amplitudes, whose two independent
linear combinations,$ $ $\delta\rho_{j,k}(t)\equiv\psi_{j}^{(0)*}c_{j,k}(t)+\mathrm{c.c.}$
and $\delta\Theta_{j,k}(t)\equiv(\psi_{j}^{(0)*}c_{j,k}(t)-\mathrm{c.c.})/i$,
are related to density and phase fluctuations, respectively. From
the deterministic part of Eq.~(\ref{eq:EOM_2c_doc}), we can directly
get the EOM for $\delta\rho_{j,k}(t)$ and $\delta\Theta_{j,k}(t)$
up to their leading order, whose explicit forms read
\begin{equation}
\partial_{t}\left(\begin{array}{c}
\delta\rho_{1,k}\\
\delta\Theta_{1,k}\\
\delta\rho_{2,k}\\
\delta\Theta_{2,k}
\end{array}\right)=\Xi\left(\begin{array}{c}
\delta\rho_{1,k}\\
\delta\Theta_{1,k}\\
\delta\rho_{2,k}\\
\delta\Theta_{2,k}
\end{array}\right),\label{eq:EOM_for_fluctuations}
\end{equation}
where $\Xi$ is a $4\times4$ matrix with the explicit form $ $
\begin{eqnarray}
 &  & \Xi=\left(\begin{array}{cc}
\Xi_{11} & \Xi_{12}\\
\Xi_{21} & \Xi_{22}
\end{array}\right),\,\Xi_{j\bar{j}}\equiv-2M_{j}^{2}\left(\begin{array}{cc}
v_{d} & 0\\
v_{c} & 0
\end{array}\right),\\
 &  & \Xi_{jj}\equiv\left(\begin{array}{cc}
-K_{j,d}k^{2}-2M_{j}^{2}u_{j,d} & K_{j,c}k^{2}\\
-K_{j,c}k^{2}-2M_{j}^{2}u_{j,c} & -K_{j,d}k^{2}
\end{array}\right).
\end{eqnarray}
Plugging the resolution 
\begin{equation}
\delta\rho_{i,k}(t)=\delta\tilde{\rho}_{i,k}e^{\omega t},\,\delta\Theta_{i,k}(t)=\delta\tilde{\Theta}_{i,k}e^{\omega t},\label{eq:resolution_fluctuation}
\end{equation}
into the EOM (\ref{eq:EOM_for_fluctuations}), we get a set of linear
equations for $\delta\tilde{\rho}_{i,k}$ and $\delta\tilde{\Theta}_{i,k}$,
i.e., $\omega\mathbf{v}=\Xi\cdot\mathbf{v}$ with $\mathbf{v}\equiv(\delta\tilde{\rho}_{1,k},\delta\tilde{\Theta}_{1,k},\delta\tilde{\rho}_{2,k},\delta\tilde{\Theta}_{2,k})^{T}$,
from which we can get the dispersion relations $\omega_{a,k}$ with
$a=1,2,3,4,$ for the four eigenmodes of the fluctuations. The expressions
for $\omega_{a,k}$ can be obtained analytically, whose forms to the
leading order in momentum $k$ read 
\begin{eqnarray}
\omega_{a,k}=\omega_{a}^{(0)}-k^{2}\frac{\mathcal{C}_{1}+\mathcal{C}_{2}\omega_{a}^{(0)}+\mathcal{C}_{2}(\omega_{a}^{(0)})^{2}}{(\omega_{a}^{(0)})^{2}-\omega_{1}^{(0)}\omega_{3}^{(0)}},\,\mathrm{for}\, a=1,3,\label{eq:omega_1_3}\\
\omega_{a,k}=-k^{2}\frac{\mathcal{C}_{1}\pm\sqrt{\mathcal{C}_{1}^{2}-4\mathcal{C}_{0}\omega_{1}^{(0)}\omega_{3}^{(0)}}}{2\omega_{1}^{(0)}\omega_{3}^{(0)}},\,\mathrm{for}\, a=2,4,\label{eq:omega_2_4}
\end{eqnarray}
where $\omega_{a}^{(0)}$ is the zero momentum part of $\omega_{a,k}$
with the explicit forms $\omega_{2,4}^{(0)}=0$ and
\begin{eqnarray}
\omega_{1,3}^{(0)}&=&-M_{1}^{2}u_{1,d}-M_{2}^{2}u_{2,d}\mp\sqrt{\left(M_{1}^{2}u_{1,d}-M_{2}^{2}u_{2,d}\right)^{2}+4M_{1}^{2}M_{2}^{2}v_{d}^{2}}.\label{eq:omega_zero_order_1_3} 
\end{eqnarray}
Here, $\mathcal{C}_{0},\mathcal{C}_{1},\mathcal{C}_{2}$ are polynomials
of $M_{j},K_{j,\kappa}u_{j,\kappa},v_{\kappa}$ with $\kappa=c,d$,
whose explicit forms are presented in~\ref{sec:Appendix_Dissipative-Phase-separation}.
We can see that $\omega_{1,k}$ and $\omega_{3,k}$ are associated
to density fluctuations which are \emph{gapped} (finite damping rate as $k\to 0$), and $\omega_{2,k}$, 
$\omega_{4,k}$ are associated to phase fluctuations which are
gapless \emph{diffusive} modes. This phenomenology is due to the absence
of particle number conservation, and is in sharp contrast to the case
of the purely coherent two-component BEC, where all the low frequency
fluctuations are \emph{gapless} \emph{sound} modes with \emph{linear}
dispersion relations~\cite{Timmermans_PRL_1998}\emph{.}

From the form of the resolution in Eq.~(\ref{eq:resolution_fluctuation})
for the fluctuations, we can see that in order for the miscible solution
$\psi_{j}^{(0)}(x,t)$ to be stable, $\omega_{a,k}<0$ is required.
At zeroth order in momentum, this requires that $\omega_{1,3}^{(0)}<0$,
which gives rise to the condition
\begin{equation}
\tilde{v}_{d}<1,\label{eq:PS_condition_zeroth_order}
\end{equation}
with $\tilde{v}_{d}\equiv v_{d}/\sqrt{u_{1,d}u_{2,d}}$ being the
rescaled dimensionless inter-component dissipative coupling, indicating
that a large enough $v_{d}$ can drive a transition to the immiscible
phase via exponentially growing gapped density fluctuations in the homogeneous state, rendering it unstable.

When the condition (\ref{eq:PS_condition_zeroth_order}) is satisfied,
requiring $\omega_{a,k}<0$ indicates the dispersion relations for
the two diffusive mode $\omega_{2,k}$, $\omega_{4,k}$ should
be both negative at the leading order in momentum, which gives rise
to the condition 
\begin{equation}
\mathcal{C}_{1}>0\,\mathrm{and}\,\mathcal{C}_{0}>0,\label{eq:PS_condition_leading_order}
\end{equation}
whose explicit form in terms of $K_{j,\kappa},u_{j,\kappa},v_{\kappa}$
can be found in appendix~\ref{sec:Appendix_Dissipative-Phase-separation}.
When $v_{d}=0$, the condition (\ref{eq:PS_condition_leading_order})
reads
\begin{eqnarray}
\fl \tilde{v}_{c}&<&\left[1+\mathcal{R}(K_{1})\mathcal{R}(u_{1})+\mathcal{R}(K_{2})\mathcal{R}(u_{2})+\mathcal{R}(K_{1})\mathcal{R}(u_{1})\mathcal{R}(K_{2})\mathcal{R}(u_{2})\right]^{1/2},
\label{eq:PS_condition_leading_order_at_zero_vd}
\end{eqnarray}
where $\tilde{v}_{c}\equiv v_{c}/\sqrt{u_{1,c}u_{2,c}}$ is the rescaled
dimensionless inter-component coherent coupling strength and function
$\mathcal{R}(z)\equiv\Re(z)/\Im(z)$ for $z\in\mathbb{C}$. This is
to be compared to the related condition in the purely coherent BEC
case, where $\tilde{v}_{c}<1$ is required to avoid the transition
to the immiscible phase~\cite{Timmermans_PRL_1998}, indicating that
particularly in the absence of a dissipative inter-component dissipative
coupling $v_{d}$, the two-component DOCs are generally more stable
against the gapless\emph{ }(diffusive) fluctuations.

Requiring both the gapped density fluctuations and the diffusive phase
fluctuations to decay exponentially with respect to time, i.e., both
condition (\ref{eq:PS_condition_zeroth_order}) and (\ref{eq:PS_condition_leading_order})
are satisfied, gives rise to the miscible-immiscible transition boundary
of the system {[}cf.~Fig.~\ref{Fig. Phase_diagram}(a) for the boundary
in terms of $\tilde{v}_{c}$ and $\tilde{v}_{d}$ at a particular
choice of other parameters{]}. Indeed, we observe in numerical simulations
that once $(\tilde{v}_{d},\tilde{v}_{c})$ is tuned outside the miscible-immiscible
transition boundary, 
despite being initialized with a generic homogeneous
configuration in the miscible phase, the system quickly evolves into
an immiscible phase, where different components occupy different spatial
regions {[}cf. Figs.~\ref{Fig. Phase_diagram}(b1) and (b2){]}. 

The stability analysis presented above is independent of dimension, and we thus expect the miscible-immiscible transition to be present in any dimension. Moreover, we expect this transition behavior persists to finite noise levels. Fig.~\ref{Fig. Phase_diagram}(b2) shows a snapshot of the system for a \emph{single} stochastic trajectory, where an immiscible phase or a phase separation can be clearly identified. However, the phase separation behavior is not expected to show in the trajectory ensemble averaged density distribution in one dimension, since the locations of the phase separated regions in different stochastic trajectories are random and the interfaces between the domains are pointlike.  As an ensemble average signature of the immiscible phase, one may expect exponential scaling of the temporal correlation function beyond the scale set by the typical size of phase separated regions, cutting off the subexponential diffusive or KPZ scaling expected at weak noise level and nonequilibrium strength (to be discussed in the following section). Moreover, position resolved density-density correlation function between different components should also reveal the signature of the immiscible phase. 

\section{Long wavelength properties of the two-component DOC in the miscible
phase}

Now let us discuss the long wavelength properties of the system. We
have seen from the previous section that there are two phases in the
system, namely, miscible and immiscible phase. In the \emph{immiscible
phase}, the different components occupy different spatial regions.
Due to the fact that all the couplings in the system are local, one
naturally expects that the two-component DOC system reduces to two
independent single-component ones with the properties revealed in
previous investigations~\cite{Altman_PRX_2015,Gladilin_PRA_2014,Ji_PRB_2015,He_PRB_2015,Wachtel_PRB_2016,Sieberber_PRB_2016,He_PRL_2017}. 

Therefore, in the following discussion, we shall only focus on the
long wavelength properties of the \emph{miscible phase}. To this end,
we first derive a low frequency effective description of the system's
dynamics in the miscible phase, and then investigate the long wavelength
scaling behavior of the system by studying the condensate phase roughness
and fluctuation functions as specified below.

\subsection{Low frequency effective description of the miscible phase}

For the single-component DOC, in the absence of phase defects, the
low frequency dynamics is effectively described by the single-component
KPZ equation~\cite{KPZ_PRL_1986} for the phase of the condensate
field~\cite{Altman_PRX_2015}. Following the lines of the derivation
presented in ~\cite{Altman_PRX_2015}, we obtain an effective description
for the low frequency dynamics for the two-component DOC, and find
it assumes the form of a two-component KPZ equation with inter-component
couplings which reads
\begin{equation}
\partial_{t}\theta_{j}=\sum_{j'=1,2}\left[D_{jj'}\partial_{x}^{2}\theta_{j'}+\lambda_{jj'}\left(\partial_{x}\theta_{j'}\right)^{2}\right]+\xi_{j},\, j=1,2,\label{eq:2-component-KPZ}
\end{equation}
where $D_{jj'},\,\lambda_{jj'},\,\sigma_{j}^{\mathrm{KPZ}}$ are rational
functions of $K_{j,\kappa},r_{j,\kappa},u_{j,\kappa},v_{\kappa}$$,\sigma_{j}$,
whose explicit forms and derivation details are presented in appendix~\ref{sec:Derivation_of_2c_KPZ}.
Here, $D_{jj'}$ characterizes phase diffusion, $\xi_{j}$ is a Gaussian
white noise field of strength $2\sigma_{j}^{\mathrm{KPZ}}$, i.e.,
$\langle\xi_{j}(x,t)\xi_{j'}(x',t')\rangle=2\sigma_{j}^{\mathrm{KPZ}}\delta_{jj'}\delta(x-x')\delta(t-t')$.
The nonlinearities $\lambda_{jj'}$ characterize the system's deviation
from thermodynamic equilibrium. Indeed, from the explicit form of $\lambda_{jj'}$
presented in appendix~\ref{sec:Derivation_of_2c_KPZ}, we can see
that all the nonlinearities $\lambda_{jj'}$ vanish identically under
detailed balance conditions, i.e., $\mathcal{R}(K_{j})=\mathcal{R}(r_{j})=\mathcal{R}(u_{j})=\mathcal{R}(v)$,
which is similar to the single-component DOC case ~\cite{Altman_PRX_2015}.

We remark here that the effective description (\ref{eq:2-component-KPZ}) is built on the assumption that the compactness of the phase fields can be neglected. As has been shown in related investigations in 1D single component DOCs~\cite{He_PRL_2017}, this assumption is a good approximation at low noise level and under weak nonequilibrium condition. However, it breaks down at high noise levels or under strong nonequilibrium conditions, where the compactness of the phase fields is expected to strongly influence the physics of the system and, therefore, has to be taken into account carefully. In these cases, we expect the low frequency effective description should assume the form of a compact version of equation (\ref{eq:2-component-KPZ}), similar to what has been shown in the single-component DOC~\cite{He_PRL_2017}. We leave the study of the system at high noise level, or under strong nonequilibrium condition, to a future investigation.

\subsection{Scaling behavior of roughness function for two-component KPZ equation
and DOC\label{sec.roughness_function}} 

Since the long wavelength dynamics of the system is effectively described
by the two-component KPZ equation (\ref{eq:2-component-KPZ}), let
us first investigate its long wavelength scaling properties. To this
end, we investigate the so-called ``roughness function'' $W_{j}(L,t)$
$ $ of the two-component KPZ equation, which is defined as $W_{j}(L,t)\equiv\langle L^{-1}\int_{x}\theta_{j}^{2}(x,t)-\left[L^{-1}\int_{x}\theta_{j}(x,t)\right]^{2}\rangle$
and measures the spatial averaged fluctuation of $\theta_{j}(x,t)$
at time $t$ of a finite system with the linear size $L$ under 
periodic boundary conditions. The importance of the roughness function
$W_{j}(L,t)$ lies in the fact that its scaling behavior with respect
to $L$ and $t$ reveals the static and dynamical critical exponents
of the system, denoted as $\alpha$ and $z$, respectively. We remark here that in order to use the roughness function as a tool to reveal the universality class of the dynamics, the phase roughness of initial states should be considerably smaller than the saturation value of the roughness function at the corresponding fixed system size. In the context of exciton-polariton condensates, this type of initial condition could possibly be achieved by imposing an additional resonant laser that depresses the spatial phase fluctuations of the condensate.

In Fig.~\ref{Fig. Phase_diagram}(c2), we show the roughness function
$W_{j}(L,t)$ for the two-component KPZ equation (\ref{eq:2-component-KPZ})
at a generic set of parameters in the scaling axes using the critical
exponents of the 1D \emph{single-component} KPZ equation, i.e., $z=3/2,\,\alpha=1/2$~\cite{KPZ_PRL_1986,Halpin-Healy_PhysRep_1995}.
We notice good finite size critical scaling collapses for both $W_{1}(L,t)$
and $W_{2}(L,t)$. This suggests that at generic choices of parameters,
the two-component KPZ equation belongs to the dynamical universality
class of the 1D\emph{ single-component} KPZ equation. 

From the scaling behavior of the two-component KPZ equation shown
in the above discussion, we naturally expect the low frequency dynamics
of the two-component DOC  [described by the full SCGLE Eq.~(\ref{eq:EOM_2c_doc})] at generic choice of parameters should also
belong to the dynamical universality class of the 1D \emph{single-component}
KPZ equation. Indeed, as we can see from Fig.~\ref{Fig. Phase_diagram}(c1),
the phase roughness functions $W_{j}(L,t)$s for the two-component
DOC indeed show good finite size critical scaling collapses by employing
the critical exponents of the 1D\emph{ single-component} KPZ equation. 

We remark here that other forms of 1D two-component KPZ equations also
emerge in different contexts in nonequilibrium statistical physics,
ranging from the early study on directed lines~\cite{Ertas_PRL_1992}
to the more recent investigations on 1D nonlinear fluctuating hydrodynamics~\cite{Ferrari_J_Stat_Phys_2013,Popkov_PRL_2014,Spohn_J_Stat_Phys_2015,Popkov_J_Stat_Phys_2015,Popkov_PNAS_2015},
where the 1D \emph{single-component} KPZ scaling behavior were found
to be ubiquitous in generic cases~\cite{Ertas_PRL_1992,Ferrari_J_Stat_Phys_2013,Spohn_J_Stat_Phys_2015,Popkov_J_Stat_Phys_2015},
despite different scaling behaviors were also found at special choices
of fine-tuned parameters in these systems~\cite{Ertas_PRL_1992,Popkov_PRL_2014,Spohn_J_Stat_Phys_2015,Popkov_J_Stat_Phys_2015,Popkov_PNAS_2015}.
We expect that the emergent 1D \emph{single-component} KPZ universal
behavior for the two-component KPZ equation Eq.(\ref{eq:2-component-KPZ})
and the dynamics of the two-component DOC at generic choices of parameters
originates from an effective decoupling at large wavelengths, which
could be further clarified via an RG analysis. Moreover, it is intriguing
to speculate that in the dynamics of two-component DOCs, nonequilibrium
scaling behavior different from the one of the 1D \emph{single-component}
KPZ equation may also arise at certain fine-tuned parameters. However,
the investigations along these lines are beyond the scope of the current
work and we leave them to a future investigation.

\subsection{Scaling behavior of spatial and temporal phase fluctuations $\Delta_{j}^{x}(x_{1},x_{2})$
and $\Delta_{j}^{t}(t_{1},t_{2})$}

Let us continue to investigate the long wavelength properties of spatial
and temporal phase fluctuation functions, i.e., $\Delta_{j}^{x}(x_{1},x_{2},t)\equiv\langle[\theta_{j}(x_{1},t)-\theta_{j}(x_{2},t)]^{2}\rangle-\langle\theta_{j}(x_{1},t)-\theta_{j}(x_{2},t)\rangle^{2}$
and $\Delta_{j}^{t}(t_{1},t_{2})\equiv L^{-1}\int_{x}\langle[\theta_{j}(x,t_{1})-\theta_{j}(x,t_{2})]^{2}\rangle-\langle\theta_{j}(x,t_{1})-\theta_{j}(x,t_{2})\rangle^{2}$,
where both quantities are measured after the system has reached its
steady state, i.e., $t,t_{1},t_{2}>T_{s}$, with $T_{s}$ equilibration
time needed for the system to reach the steady state, which is determined
by a power law with respect to the linear system size $L$, i.e.,
$T_{s}\propto L^{z}$~\cite{He_PRB_2015}. 

\begin{figure}
\centering
\includegraphics[height=1.68in]{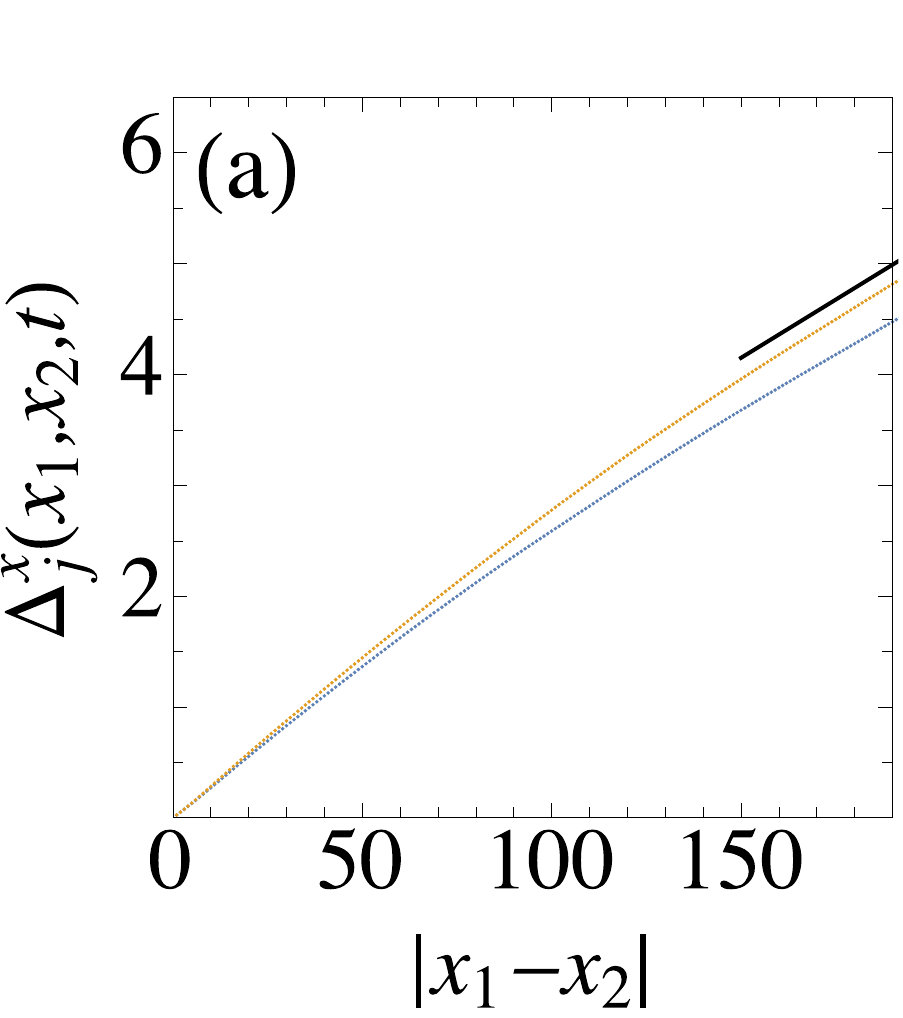}$\,\,\,\,\,\,\,$\includegraphics[height=1.6in]{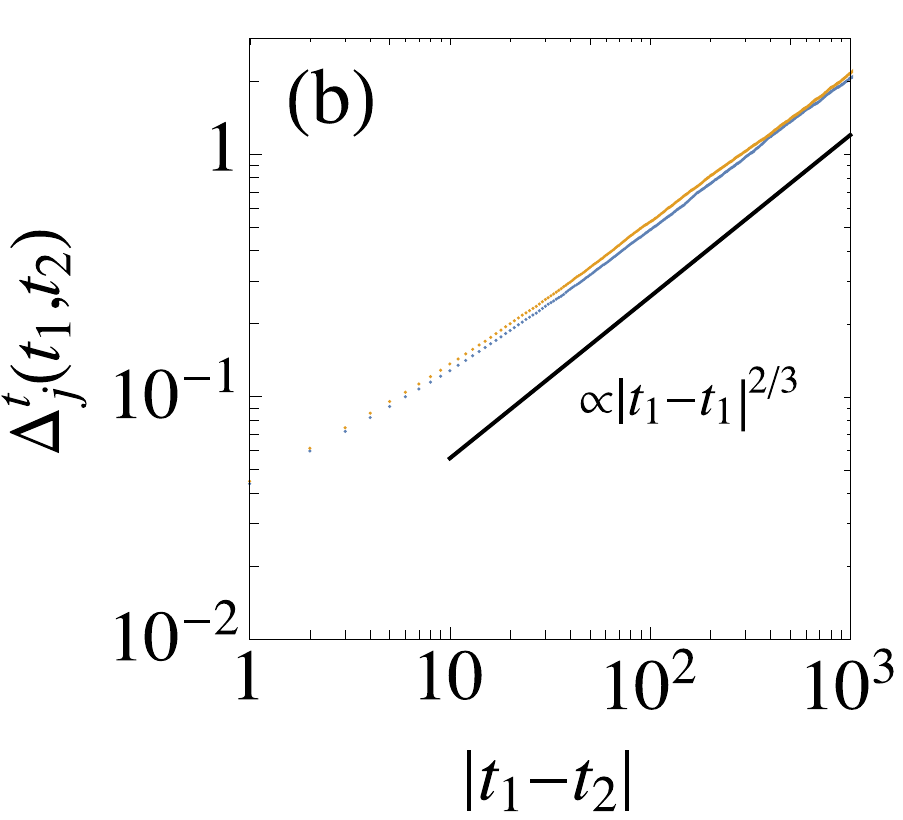}

\caption{Spatial and temporal phase fluctuation functions $\Delta_{j}^{x}(x_{1},x_{2},t)$
and $\Delta_{j}^{t}(t_{1},t_{2})$ for the same set of parameters
as the one in Fig.~\ref{Fig. Phase_diagram}(c1) with system size
$L=2^{10}$. (a) Scaling behavior of spatial phase fluctuation function
$\Delta_{j}^{x}(x_{1},x_{2},t)$ on a linear scale. The black solid
line corresponds to a linear function of $|x_{1}-x_{2}|$. The lower
and upper curve correspond to $\Delta_{1}^{x}(x_{1},x_{2},t)$ and
$\Delta_{2}^{x}(x_{1},x_{2},t)$, respectively, both of which show
linear behavior at relatively large distances. (b) Scaling behavior
of the temporal phase fluctuation function $\Delta_{j}^{t}(t_{1},t_{2})$
on a double logarithmic scale. The black solid line corresponds to
a power law $\propto|t_{1}-t_{2}|^{2/3}$. The lower and upper curves
correspond to $\Delta_{1}^{t}(t_{1},t_{2})$ and $\Delta_{2}^{t}(t_{1},t_{2})$,
respectively, where one can observe the power law behavior at
large time differences. Both the spatial and temporal
phase fluctuation functions show scaling behavior that is consistent
with the universality class of the 1D \emph{single-component} KPZ
equation. See text for more details.}

\label{Fig.phase_fluc} 
\end{figure}

Fig.~\ref{Fig.phase_fluc} shows the spatial and temporal phase fluctuation
functions $\Delta_{j}^{x}(x_{1},x_{2},t)$ and $\Delta_{j}^{t}(t_{1},t_{2})$
for the same set of parameters as the one for Fig.~\ref{Fig. Phase_diagram}(c1).
We notice that the spatial phase fluctuation functions $\Delta_{j}^{x}(x_{1},x_{2},t)$
show a linear growth at relatively large distances, i.e., $\Delta_{j}^{x}(x_{1},x_{2},t)\propto|x_{1}-x_{2}|$,
while the temporal phase fluctuation functions $\Delta_{j}^{t}(t_{1},t_{2})$
show a power law growth with an exponent $2/3$ at relatively large
time differences, i.e., $\Delta_{j}^{t}(t_{1},t_{2})\propto|t_{1}-t_{2}|^{2/3}$.
Again, these observations are consistent with the expectations from
the universality class of the 1D \emph{single-component} KPZ equation,
where the scaling behavior of $\Delta_{j}^{x}(x_{1,}x_{2},t)$ and
$\Delta_{j}^{t}(t_{1},t_{2})$ are expected to be determined by the
static exponent $\alpha$ and the so-called growth exponent $\beta$,
respectively, i.e., $\Delta_{j}^{x}(x_{1,}x_{2},t)\propto|x_{1}-x_{2}|^{2\alpha}$
and $\Delta_{j}^{t}(t_{1},t_{2})\propto|t_{1}-t_{2}|^{2\beta}$, with
$\alpha=1/2$ and $\beta=1/3$ for the 1D\emph{ single-component}
KPZ equation~\cite{KPZ_PRL_1986,Halpin-Healy_PhysRep_1995}.

\section{Experimental observability}

\begin{figure}
\centering
\includegraphics[height=1.5in]{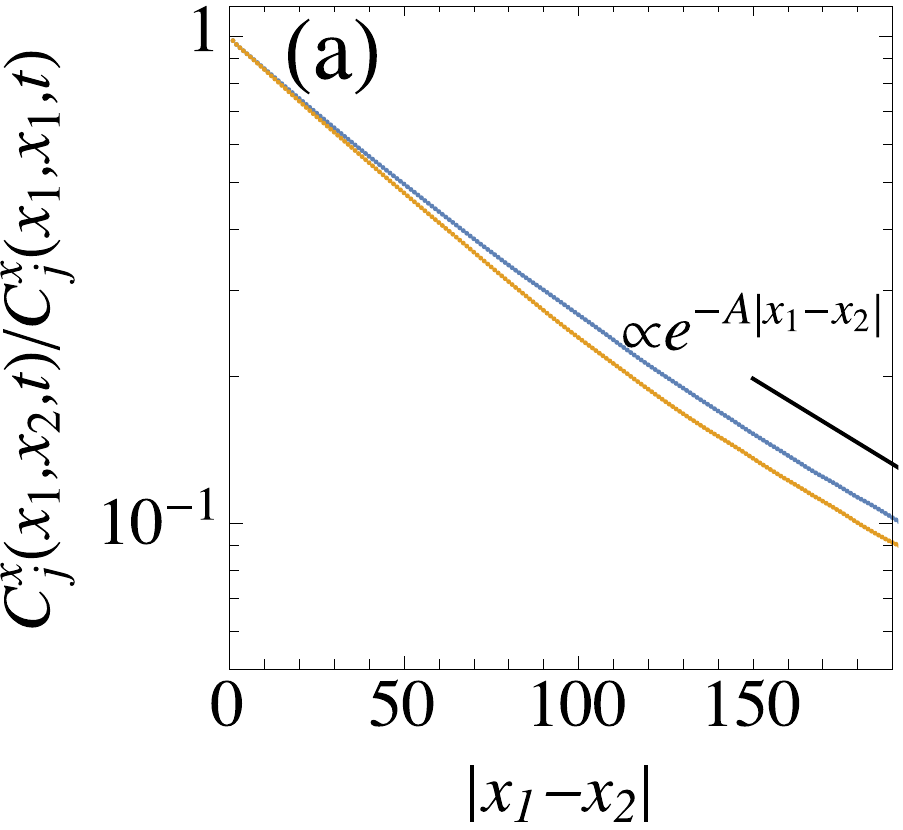}$\,\,\,\,\,\,\,$\includegraphics[height=1.5in]{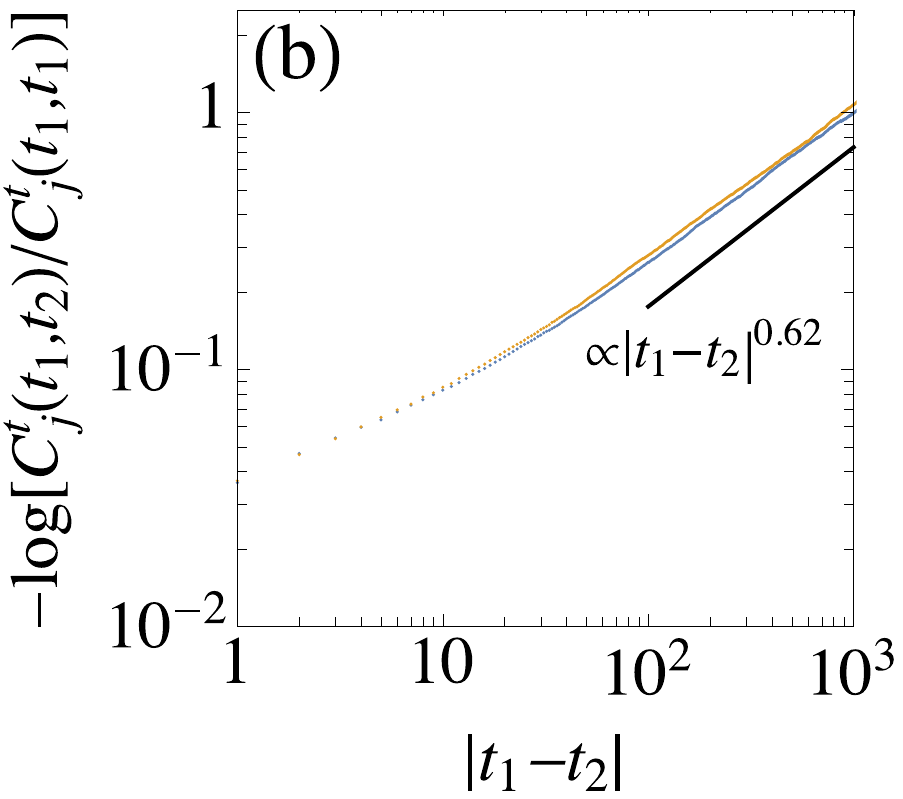}

\caption{The decay behavior of the spatial and temporal correlation functions
$C_{j}^{x}(x_{1},x_{2},t)$ and $C_{j}^{t}(t_{1},t_{2})$ for the
same set of parameters as the one in Fig.~\ref{Fig. Phase_diagram}(c1)
with system size $L=2^{10}$. (a) Spatial correlation function $C_{j}^{x}(x_{1},x_{2},t)$
on a semi-logarithmic scale. The black solid line corresponds to a
exponential function $\propto e^{-A|x_{1}-x_{2}|}$. The upper and
lower curve correspond to $C_{1}^{x}(x_{1},x_{2},t)$ and $C_{2}^{x}(x_{1},x_{2},t)$,
respectively, both of which show an exponential decay at relatively
large distances. (b) The dependence of $-\log[C_{j}^{t}(t_{1},t_{2})/C_{j}^{t}(t_{1},t_{1})]$
on $|t_{1}-t_{2}|$ on a double-logarithmic scale. The black solid
line corresponds to a power law function $\propto|t_{1}-t_{2}|^{0.62}$.
The lower and upper curve correspond to $C_{1}^{t}(t_{1},t_{2})$
and $C_{2}^{t}(t_{1},t_{2})$, respectively, both of which show the
sub-exponential decay behavior at relatively large time differences.
A linear fit to the data points with $|t_{1}-t_{2}|\in[10^{2},10^{3}]$
gives rise to $\beta=0.31$. We observe that the decay behavior of
both spatial and temporal correlation functions are consistent with
the universality class of the 1D\emph{ single-component} KPZ equation.}

\label{Fig.corr} 
\end{figure}

For the KPZ physics of the two-component DOC in its miscible phase,
we expect the characteristic signatures are observable in the two-point
spatial and temporal correlation functions $C_{j}^{x}(x_{1},x_{2},t)\equiv\langle\psi_{j}^{*}(x_{1},t)\psi_{j}(x_{2},t)\rangle$
and $C_{j}^{t}(t_{1},t_{2})\equiv L^{-1}\int_{x}\langle\psi_{j}^{*}(x,t_{1})\psi_{j}(x,t_{2})\rangle$
in the system's steady state, i.e., $t,t_{1},t_{2}>T_{s}$, both of
which are accessible in exciton-polariton experiments~\cite{Kasprzak_nature_2006,Roumpos_PNAS_2012,Love_PRL_2008}.
More specifically, from the long wavelength scaling behavior of the
two-component DOC presented previously, we expect $C_{j}^{x}(x_{1},x_{2},t)\propto e^{-A_j|x_{1}-x_{2}|^{2\alpha}}$
and $C_{j}^{t}(t_{1},t_{2})\propto e^{-B_j|t_{1}-t_{2}|^{2\beta}}$,
where $\alpha=1/2$, $\beta=1/3$ are the static and growth exponent
for the 1D \emph{single-component} KPZ equation~\cite{KPZ_PRL_1986,Halpin-Healy_PhysRep_1995},
respectively, and $A_j,B_j$ are two nonuniversal positive constants determined
by the microscopic details of the system. Fig.~\ref{Fig.corr} shows
the decay behavior of the spatial and temporal correlation function
$C_{j}^{x}(x_{1},x_{2},t)$ and $C_{j}^{t}(t_{1},t_{2})$ for the
same set of parameters as the one for Fig.~\ref{Fig. Phase_diagram}(c1).
Indeed, we observe that the decay behavior of both spatial and temporal
correlation function are consistent with the expectations from the
universality class of the KPZ equation.

Our discussion is based on a generic model for two-component DOC,
hence we expect it is relevant for different experimental realizations.
Currently, the most promising realization could come from the two-component
polariton condensate systems, where the two different components correspond
to the two different polarization directions of the photonic part
of the polaritons~\cite{Takemura_NatPhys_2014,Fischer_PRL_2014,Ohadi_PRX_2015,Askitopoulos_PRB_2016}.
We can see from Fig.~\ref{Fig. Phase_diagram}(a) that both the coherent and the dissipative inter-component interaction can drive the miscible-immiscible transition.
In particular, the recently achieved tunability of inter-component
coupling strengths via the polaritonic Feshbach resonance~\cite{Takemura_NatPhys_2014}
gives rise to an intriguing platform for experimental investigations
on the miscible-immiscible transition phenomenon in two-component
DOCs established in this work. We expect that single experimental runs should indeed reveal the fragmentation of the driven open condensate in real space, leaving a clear fingerprint of the immiscible phase in one dimension. Moreover, the scaling of the temporal coherence function is expected to exhibit a difference from the one of the miscible phase, which manifests itself beyond the new scale set by the typical size of the phase separated domains.

Finally, we remark that compared to current typical setups in experiments
for two-component polariton condensates ~\cite{Takemura_NatPhys_2014,Fischer_PRL_2014,Ohadi_PRX_2015,Askitopoulos_PRB_2016}, for instance, those investigating the polarization dynamics of polariton condensates~\cite{Fischer_PRL_2014,Ohadi_PRX_2015,Askitopoulos_PRB_2016},
the model for two-component DOCs studied here possesses a higher symmetry,
namely an invariance under independent phase rotations for different
components, i.e., $\psi_{j}\rightarrow\psi_{j}e^{i\phi_{j}}$. The presence of this symmetry warrants the 
interesting physical scenarios discussed
previously. In particular, it gives rise to a more interesting two-component KPZ equation for the miscible phase that possibly hosts nonequilibrium dynamical scaling behavior beyond the one of the single-component KPZ equation as shortly mentioned in Sec.~\ref{sec.roughness_function}.  Although there exists rich physics even in the absence of the independent phase rotation symmetry, in both purely coherent \cite{Nicklas_PRL_2015} and driven-open \cite{Ohadi_PRX_2015} two-component condensates, in order to experimentally investigate the physics discussed in current work, one should reduce and ideally completely eliminate the strength of factors in experiments that can cause the breaking of independent phase rotation symmetry. For instance, asymmetry at the interfaces of quantum-wells, or the anisotropy-induced splitting of linear polarizations in the microcavity~\cite{Ohadi_PRX_2015, Askitopoulos_PRB_2016} should be kept as small as possible, so that the physical effects originating from the weak breaking of the independent phase rotation symmetry are negligible at the typical spatio-temporal scale in experiments. 

In fact for the KPZ physics,
we do not expect observable modifications when independent phase rotation
symmetry is absent. This is due to the fact that the effective description
of the system in this case is expected to be directly described by
the single-component KPZ equation that corresponds to the identical
phase rotation symmetry for both components, i.e., $\psi_{j}\rightarrow\psi_{j}e^{i\phi}$.
To check the robustness of the miscible-immiscible transition against symmetry breaking perturbations, we checked a concrete case where the system is exposed to a weak single-particle inter-component exchange process whose strength is around $5\%$ of $r_{j,c}$ in the case of Figs.~\ref{Fig. Phase_diagram}(b1,b2). We found the misible-immiscible transition phenomenon is not substantially changed (see~\ref{sec:Effects-of-weak-symmetry-breaking} for more details).

\section{Conclusions and outlook}

Two-component DOCs give rise to a novel scenario for the miscible-immiscible transition 
compared to its purely closed system counterpart, since
the dynamical resources that cause the transition are qualitatively
changed due to the absence of particle number and momentum
conservation. In particular, the transition is driven predominantly
by the gapped density fluctuations generated via the inter-component
dissipative coupling. Below the transition in the miscible phase, we find the low frequency dynamics of the system to be effectively
described by a two-component KPZ equation, which however belongs to
the same nonequilibrium universality class as the single-component
KPZ equation in 1D at generic choices of parameters. We believe that
our work will stimulate further theoretical and experimental investigations
on multicomponent DOCs. On the experimental side, the observation
of the miscible-immiscible transition in a driven-open context
in multicomponent polariton systems would complement the closed system
counterpart observed in ultracold atoms~\cite{Myatt_PRL_1997,Hall_PRL_1998,Nicklas_PRL_2011}
in a fundamentally different physical context, underpinning the generality
of this phenomenon. On the theoretical side, the investigation of
the high noise level and strong nonequilibrium regimes appears most
interesting, where the phase compactness as an ingredient beyond the
usual KPZ scenario must be expected to play a crucial role.

\ack
We thank the Center for Information Services and High Performance
Computing (ZIH) at TU Dresden for allocation of computer time. This
work was supported by German Research Foundation (DFG) through ZUK
64, through the Institutional Strategy of the University of Cologne
within the German Excellence Initiative (ZUK 81) and by the European
Research Council via ERC Grant Agreement No. 647434 (DOQS).

\appendix

\section{Miscible-immiscible transition\label{sec:Appendix_Dissipative-Phase-separation}}

In this appendix, we present some calculation details involved in
the discussion for the miscible-immiscible transition. 

It is easy to notice that $r_{j,c}$ can be chosen in such a way that
there exists a stationary, spatially uniform solution for the two-component
CGLE, i.e., the deterministic part of the two-component SCGLE in Eq.~(\ref{eq:EOM_2c_doc}),
\begin{eqnarray}
\partial_{t}\psi_{j} & = & \left(K_{j}\partial_{x}^{2}+r_{j}-u_{j}|\psi_{j}|^{2}-v|\psi_{\bar{j}}|^{2}\right)\psi_{j},
\end{eqnarray}
which reads $\psi_{j}^{(0)}(x,t)=M_{j}$ as shown in the main text.
Here, $M_{j}$ is the solution for the real homogenous part of the
two-component CGLE, i.e., $r_{j,d}-u_{j,d}M_{j}^{2}-v_{d}M_{\bar{j}}^{2}=0,$
whose explicit form reads $M_{j}^{2}=(r_{j,d}u_{\bar{j},d}-r_{\bar{j},d}v_{d})/(u_{1,d}u_{2,d}-v_{d}^{2})$.
For a generic choice of parameters, the existence of the solution
for $M_{1}^{2}$ and $M_{2}^{2}$ requires 
\begin{equation}
\det\left(\begin{array}{cc}
u_{1.d} & v_{d}\\
v_{d} & u_{2,d}
\end{array}\right)=u_{1,d}u_{2,d}-v_{d}^{2}\neq0.
\end{equation}
We notice that there is an additional constraint on the parameters
originating from $M_{j}^{2}>0$, which gives rise to
\begin{equation}
\fl
\frac{v_{d}}{u_{2,d}}>\frac{r_{1,d}}{r_{2,d}}>\frac{u_{1,d}}{v_{d}},\,\mathrm{if}\, u_{1,d}u_{2,d}-v_{d}^{2}>0,\,\mathrm{and}\,\frac{v_{d}}{u_{2,d}}<\frac{r_{1,d}}{r_{2,d}}<\frac{u_{1,d}}{v_{d}},\,\mathrm{if}\, u_{1,d}u_{2,d}-v_{d}^{2}<0.\label{eq:positive_M1sqr_M2sqr_condition_stable_case}
\end{equation}
The above stationary, spatially uniform solution is not always stable
against small fluctuations. In the following, we perform a leading
order stability analysis following a similar approach that has been
applied to the coherent two-component BEC in Ref.~\cite{Timmermans_PRL_1998}.
As we have seen in the main text, the dispersion relation of eigenmodes
$\omega_{a,k}$ up to the leading order in momentum can be expressed
in a compact form by the help of three polynomials, $\mathcal{C}_{0},\mathcal{C}_{1},$
and $\mathcal{C}_{2}$ in terms of $M_{j},K_{j,\kappa}u_{j,\kappa},v_{\kappa}$,
whose explicit forms read
\begin{eqnarray}
\fl
\mathcal{C}_{0}  = 4M_{1}^{2}M_{2}^{2}\left[K_{1,c}\left(u_{1,c}K_{2,d}u_{2,d}-v_{c}v_{d}K_{2,d}+K_{2,c}u_{1,c}u_{2,c}-v_{c}^{2}K_{2,c}\right)\right.\\
\fl 
\left.+K_{1,d}\left(K_{2,c}u_{2,c}u_{1,d}-v_{c}v_{d}K_{2,c}+K_{2,d}u_{1,d}u_{2,d}-v_{d}^{2}K_{2,d}\right)\right],\nonumber \\
\fl
\mathcal{C}_{1} =4M_{1}^{2}M_{2}^{2}\left[K_{2,c}u_{2,c}u_{1,d}+K_{1,c}u_{1,c}u_{2,d}-v_{c}v_{d}\left(K_{1,c}+K_{2,c}\right)+K_{1,d}u_{1,d}u_{2,d}\right.\\
\fl
\left.+K_{2,d}u_{1,d}u_{2,d}-v_{d}^{2}\left(K_{1,d}+K_{2,d}\right)\right],\nonumber\\
\fl
\mathcal{C}_{2}  =  2\left[M_{1}^{2}K_{1,c}u_{1,c}+M_{2}^{2}K_{2,c}u_{2,c}+M_{1}^{2}\left(K_{1,d}+2K_{2,d}\right)u_{1,d}+M_{2}^{2}\left(2K_{1,d}+K_{2,d}\right)u_{2,d}\right]. \nonumber\\
\end{eqnarray}
Requiring $\omega_{2,4}<0$ gives rise to the miscible condition $\mathcal{C}_{1}>0$
and simultaneously $\mathcal{C}_{0}>0$, whose explicit form reads
\begin{eqnarray}
 \fl \tilde{v}_{c}<\frac{\sqrt{\mathcal{R}(u_{1})\mathcal{R}(u_{2})}}{\tilde{v}_{d}}\left[\mathcal{R}(K_{1}+K_{2})+\frac{1}{(K_{1,c}+K_{2,c})}\left(\frac{K_{2,c}}{\mathcal{R}(u_{2})}+\frac{K_{1,c}}{\mathcal{R}(u_{1})}\right)\right]\nonumber \\
 \fl \mathrm{and\, simultaneously}\nonumber \\
 \fl \tilde{v}_{c}<\left\{ 1+\mathcal{R}(K_{1})\mathcal{R}(u_{1})+\mathcal{R}(K_{2})\mathcal{R}(u_{2})+\mathcal{R}(K_{1})\mathcal{R}(u_{1})\mathcal{R}(K_{2})\mathcal{R}(u_{2})\cdot(1-\tilde{v}_{d}^{2})\right.\\
\fl \left.+\left[\frac{1}{2}\sqrt{\mathcal{R}(u_{1})\mathcal{R}(u_{2})}\left(\mathcal{R}(K_{1})+\mathcal{R}(K_{2})\right)\tilde{v}_{d}\right]^{2}\right\} ^{1/2} -\frac{1}{2}\sqrt{\mathcal{R}(u_{1})\mathcal{R}(u_{2})}\left(\mathcal{R}(K_{1})+\mathcal{R}(K_{2})\right)\tilde{v}_{d},\nonumber
\end{eqnarray}
where $\tilde{v}_{c}\equiv v_{c}/\sqrt{u_{1,c}u_{2,c}}$ and $\tilde{v}_{d}\equiv v_{d}/\sqrt{u_{1,d}u_{2,d}}$
are the rescaled dimensionless inter-component coherent and dissipative
coupling strength, respectively, and the definition $\mathcal{R}(z)\equiv\Re(z)/\Im(z)$
for $z\in\mathbb{C}$. Notice that $ $the two conditions $\mathcal{C}_{1}>0$
and $\mathcal{C}_{0}>0$ are independent, therefore one should generically
solve these two inequalities and take the overlap region indicated
by the two. 

When $r_{1,d}=r_{2,d}\equiv r_{d},u_{1,d}=u_{2,d}\equiv u_{d},v_{d}=0$,
$\omega_{1}^{(0)}$ and $\omega_{3}^{(0)}$ are degenerate, and the
dispersion relation assumes the following form
\begin{eqnarray}
\fl
\omega_{a,k}=-2M^{2}u_{d}-\frac{k^{2}}{2u_{d}}\left\{ \pm\left(-K_{1,c}u_{1,c}-K_{2,c}u_{2,c}+u_{d}K_{1,d}+u_{d}K_{2,d}\right)\right.\\
\fl
+\left[-2K_{1,c}\left(u_{d}u_{1,c}K_{1,d}-u_{d}u_{1,c}K_{2,d}+K_{2,c}u_{1,c}u_{2,c}-2v_{c}^{2}K_{2,c}\right)\right.\nonumber\\
\fl
\left.\left.+\left(K_{2,c}u_{2,c}+u_{d}K_{1,d}-u_{d}K_{2,d}\right){}^{2}+K_{1,c}^{2}u_{1,c}^{2}\right]^{1/2}\right\}, \,\mathrm{for}\, a=1,3,\nonumber\\
\fl
\omega_{a,k}=-\frac{k^{2}}{2u_{d}}\left\{ \left(K_{1,c}u_{1,c}+K_{2,c}u_{2,c}+u_{d}K_{1,d}+u_{d}K_{2,d}\right)\right.\\
\fl
\pm\left[-2K_{1,c}\left(-u_{d}u_{1,c}K_{1,d}+u_{d}u_{1,c}K_{2,d}+K_{2,c}u_{1,c}u_{2,c}-2v_{c}^{2}K_{2,c}\right)\right.\nonumber\\
\fl
\left.\left.+\left(K_{2,c}u_{2,c}-u_{d}K_{1,d}+u_{d}K_{2,d}\right){}^{2}+K_{1,c}^{2}u_{1,c}^{2}\right]^{1/2}\right\} ,\,\mathrm{for}\, a=2,4,\nonumber
\end{eqnarray}
where $M\equiv r_{d}/u_{d}$.

\section{Derivation of the two-component KPZ equation as the effective description
for two-component DOCs\label{sec:Derivation_of_2c_KPZ}}

In this appendix, we present the derivation details for the low frequency
effective description of two-component DOCs, which can be straightforwardly
applied to the generic multi-component case.

From the equation of motion (EOM) for DOCs, we can directly write
down the EOM for the real and imaginary part of $\psi_{j}$, denoted
as $\psi_{j,a}$, with $a=1,2$, i.e., $\psi_{j}=\psi_{j,1}+i\psi_{j,2}$.
The generic form of the dynamical equation for $\psi_{j,a}$ reads
\begin{equation}
\partial_{t}\psi_{j,a}=F_{j,a}[\{\psi_{j,a}\}].\label{eq:EOM_psi_ja}
\end{equation}
In order to derive a set of dynamical equations for $\rho_{j},\theta_{j}$,
where $\psi_{j,1}=\rho_{j}\cos\theta_{j}$, $\psi_{j,2}=\rho_{j}\sin\theta_{j}$,
we can make use of the Jacobian matrix between $\{\psi_{j,a}\}$ and
$\{\rho_{j},\theta_{j}\}$, i.e., 
\begin{eqnarray}
\fl
\left(\begin{array}{c}
\partial_{t}\rho_{1}\\
\partial_{t}\theta_{1}\\
\partial_{t}\rho_{2}\\
\partial_{t}\theta_{2}
\end{array}\right) & = & \left(\begin{array}{cccc}
\cos\theta_{1} & -\rho_{1}\sin\theta_{1} & 0 & 0\\
\sin\theta_{1} & \rho_{1}\cos\theta_{1} & 0 & 0\\
0 & 0 & \cos\theta_{2} & -\rho_{2}\sin\theta_{2}\\
0 & 0 & \sin\theta_{2} & \rho_{2}\cos\theta_{2}
\end{array}\right)^{-1}\left(\begin{array}{c}
\partial_{t}\psi_{1,1}\\
\partial_{t}\psi_{1,2}\\
\partial_{t}\psi_{2,1}\\
\partial_{t}\psi_{2,2}
\end{array}\right).
\end{eqnarray}
After plugging the EOM for $\psi_{j,a}$ Eq.~(\ref{eq:EOM_psi_ja})
into the above equation, and substituting $\psi_{i,1}=\rho_{i}\cos\theta_{i}$,
$\psi_{i,2}=\rho_{i}\sin\theta_{i}$, we arrive at the EOM for the
amplitude and the phase fields, i.e., 
\begin{eqnarray}
\fl
\left(\begin{array}{c}
\partial_{t}\rho_{1}\\
\partial_{t}\theta_{1}\\
\partial_{t}\rho_{2}\\
\partial_{t}\theta_{2}
\end{array}\right)=\left(\begin{array}{cccc}
\cos\theta_{1} & \sin\theta_{1} & 0 & 0\\
-\rho_{1}^{-1}\sin\theta_{1} & \rho_{1}^{-1}\cos\theta_{1} & 0 & 0\\
0 & 0 & \cos\theta_{2} & \sin\theta_{2}\\
0 & 0 & -\rho_{2}^{-1}\sin\theta_{2} & \rho_{2}^{-1}\cos\theta_{2}
\end{array}\right)\left(\begin{array}{c}
F_{1,1}[\{\rho_{j},\theta_{j}\}]\\
F_{1,2}[\{\rho_{j},\theta_{j}\}]\\
F_{2,1}[\{\rho_{j},\theta_{j}\}]\\
F_{2,2}[\{\rho_{j},\theta_{j}\}]
\end{array}\right).\nonumber\\
\label{eq:EOM_for_rho_theta}
\end{eqnarray}
The following steps in the derivation go along the lines of the similar
discussion presented in Ref.~\cite{Altman_PRX_2015} as we outline
below. We first decompose the amplitude field $\rho_{j}(x,t)$ into
the sum of the stationary, spatially uniform amplitude $|\psi_{j}^{(0)}(x,t)|=M_{j}$
and the amplitude fluctuation $\chi_{j}(x,t)$ on its top, i.e., $\rho_{j}(x,t)=M_{j}+\chi_{j}(x,t)$.
After substituting this decomposition into Eq.~(\ref{eq:EOM_for_rho_theta}),
we arrive at a EOM for amplitude and phase fluctuations. Since the
dynamics of the gapped amplitude fluctuation $\chi_{j}(x,t)$ is fast
compared to the gapless phase fluctuations, we can further adiabatically
eliminate $\chi_{j}(x,t)$ and arrive at the two-component KPZ equation
Eq.~(\ref{eq:2-component-KPZ}) for the phase fields $\theta_{i}(x,t)$,
upon keeping only the terms that are not irrelevant in the RG sense.
The explicit forms for the parameters in the two-component KPZ equation
Eq.~(\ref{eq:2-component-KPZ}) read
\begin{eqnarray}
\fl D_{11}&=&\mathcal{A}\left(K_{1,c}u_{1,c}u_{2,d}+K_{1,d}u_{1,d}u_{2d}-K_{1,c}v_{c}v_{d}-K_{1,d}v_{d}^{2}\right),\\
\fl D_{22}&=&\mathcal{A}\left(K_{2,c}u_{2,c}u_{1,d}+K_{2,d}u_{2,d}u_{1,d}-K_{2,c}v_{c}v_{d}-K_{2,d}v_{d}^{2}\right),\\
\fl D_{12}&=&\mathcal{A}K_{2,c}\left(u_{1,d}v_{c}-u_{1,c}v_{d}\right),\\
\fl D_{21}&=&\mathcal{A}K_{1,c}\left(u_{2,d}v_{c}-u_{2,c}v_{d}\right),\\
\fl \lambda_{11}&=&\left(K_{1,d}u_{1,c}u_{2,d}-K_{1,c}u_{1,d}u_{2d}-K_{1,d}v_{c}v_{d}+K_{1,c}v_{d}^{2}\right),\\
\fl \lambda_{22}&=&\left(K_{2,d}u_{2,c}u_{1,d}-K_{2,c}u_{2,d}u_{1d}-K_{2,d}v_{c}v_{d}+K_{2,c}v_{d}^{2}\right),\\
\fl \lambda_{12}&=&K_{2,d}\left(u_{1,d}v_{c}-u_{1,c}v_{d}\right),\\
\fl \lambda_{21}&=&K_{1,d}\left(u_{2,d}v_{c}-u_{2,c}v_{d}\right),\\
\fl \sigma_{1}^{\mathrm{KPZ}}&=&\frac{1}{M_{1}^{2}}\left(1+\mathcal{\mathcal{A}}^{2}\left(u_{1,c}u_{2,d}-v_{c}v_{d}\right)^{2}\right)\sigma_{1}+\frac{1}{M_{2}^{2}}\mathcal{A}^{2}\left(u_{1,d}v_{c}-u_{1,c}v_{d}\right)^{2}\sigma_{2},\\
\fl \sigma_{2}^{\mathrm{KPZ}}&=&\frac{1}{M_{2}^{2}}\left(1+\mathcal{A}^{2}\left(u_{2,c}u_{1,d}-v_{c}v_{d}\right)^{2}\right)\sigma_{2}+\frac{1}{M_{1}^{2}}\mathcal{A}^{2}\left(u_{2,d}v_{c}-u_{2,c}v_{d}\right)^{2}\sigma_{1},
\end{eqnarray}
with $\mathcal{A}\equiv\left(u_{1,d}u_{2,d}-v_{d}^{2}\right)^{-1}$.

\section{Effects of weak breaking of independent phase rotation symmetry\label{sec:Effects-of-weak-symmetry-breaking}}

As mentioned in the main text, in current experimental setups for
two-component polariton condensates~\cite{Fischer_PRL_2014,Takemura_NatPhys_2014,Ohadi_PRX_2015,Askitopoulos_PRB_2016},
there exist physical processes that can break the independent phase
rotation symmetry. One type of these processes is the single particle
inter-component exchange~\cite{Ohadi_PRX_2015,Askitopoulos_PRB_2016}
that corresponds to a term $\varepsilon\psi_{\bar{j}}$ appearing
in the right hand side of the dynamical equation for $\psi_{j}$ Eq.~(\ref{eq:EOM_2c_doc}),
i.e., the EOM of the system in the presence the single particle inter-component
exchange reads 
\begin{equation}
\partial_{t}\psi_{j}=\left(K_{j}\partial_{x}^{2}+r_{j}-u_{j}|\psi_{j}|^{2}-v|\psi_{\bar{j}}|^{2}\right)\psi_{j}+\varepsilon\psi_{\bar{j}}+\zeta_{j}.\label{eq:EOM_with_inter-component_exchange}
\end{equation}
Here $\varepsilon$ is a complex number, i.e., $\varepsilon=\varepsilon_{d}+i\varepsilon_{c}$,
whose real and imaginary part characterize the dissipative and coherent
inter-component exchange rate, respectively. Its magnitude $|\varepsilon|$
directly gives rise to a time scale $T_{B}\propto|\varepsilon|^{-1}$
and a length scale $L_{B}\propto|\varepsilon|^{-1/z}$(with $z$ being
the dynamical exponent of the system$ $), beyond which the physical
effects associated to the breaking of the symmetry are expected to
manifest themselves. This indicates that as long as $|\varepsilon|$
is small enough such that the associated spatial (temporal) scale
$L_{B}$ ($T_{B}$) is larger than the spatial (temporal) scale beyond
which the physical phenomena predicted by the theory \emph{with the
independent phase rotation symmetry} could appear, the corresponding
physical phenomena are expected not to be substantially affected. 

\begin{figure}
\centering
\includegraphics[height=1.5in]{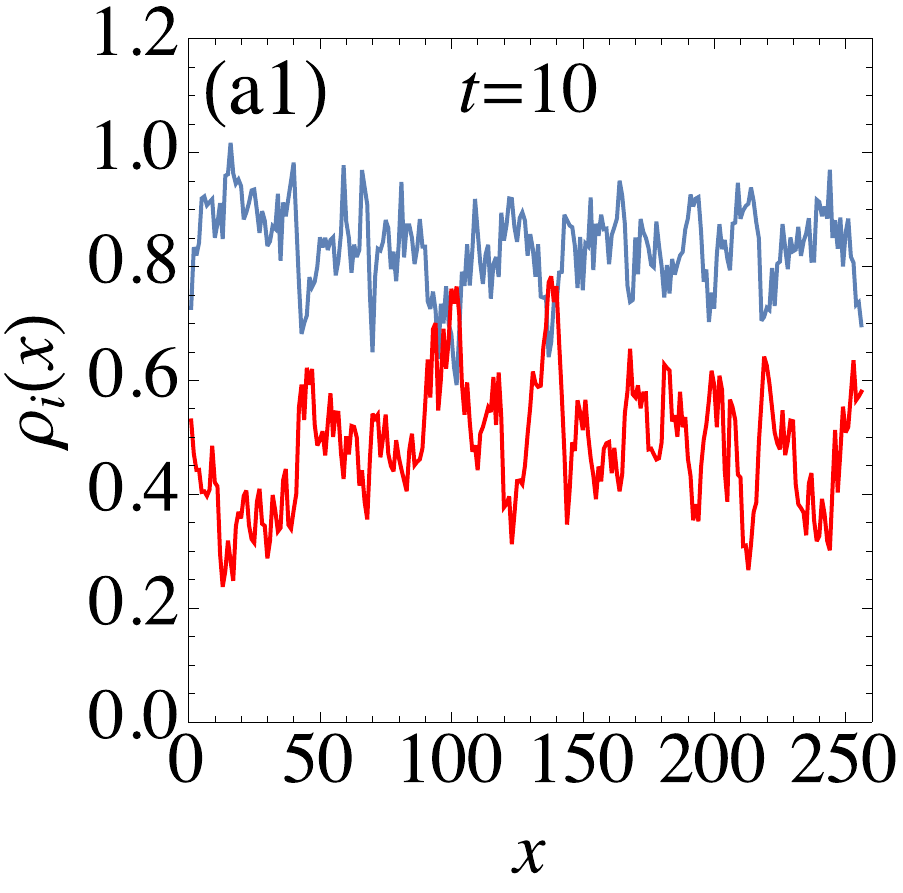}\includegraphics[height=1.5in]{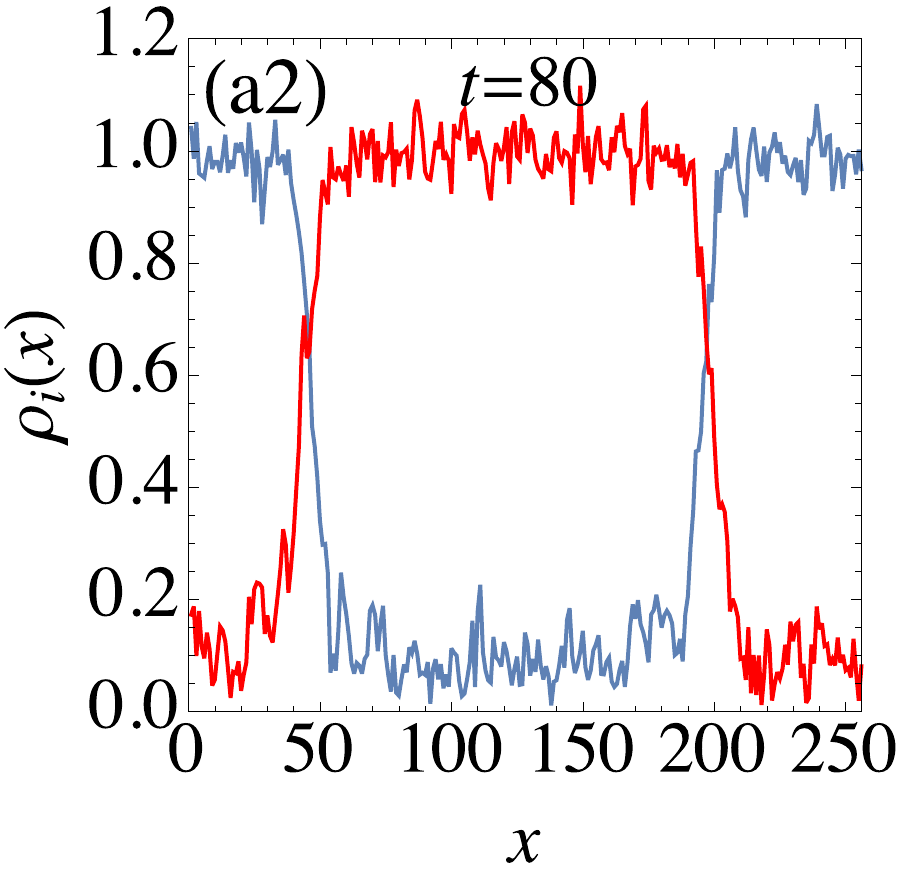}
\caption{Snapshots of distributions of condensate field amplitudes $\rho_{j}(x)\equiv|\psi_{j}(x)|$
at different time (the blue and red curve correspond to $\rho_{1}(x)$
and $\rho_{2}(x)$, respectively) of a system in the presence of a
weak inter-component exchange term with $\varepsilon_{d}=\varepsilon_{c}=0.003$,
where an immiscible phase is clearly observed at a later time ($t=80$).
The values of all other parameters are the same as those in Figs.~\ref{Fig. Phase_diagram}(b1,b2).
See text for more details.}

\label{Fig. MIM_ransition_in_presence_of_weak_symmetry_breaking} 
\end{figure}

As a concrete example, in Fig.~\ref{Fig. MIM_ransition_in_presence_of_weak_symmetry_breaking},
we show two snapshots of the density distribution at different time
of a system in the presence of weak inter-component exchange processes.
Except $|\varepsilon|$ being around $5\%$ of $r_{j,c}$, all other
parameters corresponding to Fig.~\ref{Fig. MIM_ransition_in_presence_of_weak_symmetry_breaking}
are exactly the same as those in Figs.~\ref{Fig. Phase_diagram}(b1,b2),
i.e., the system is expected to evolve into an immiscible phase despite
being initialized with a generic homogeneous configuration in the
miscible phase. As we can see from Fig.~\ref{Fig. MIM_ransition_in_presence_of_weak_symmetry_breaking},
in the presence of small $|\varepsilon|$, the system still show a
similar evolution as the case with $|\varepsilon|=0$ {[}cf.~Figs.~\ref{Fig. Phase_diagram}(b1,b2){]},
indicating that the effects of weak symmetry breaking term are not substantial
in this case.

\end{document}